\newcommand{\dgr}{{^\circ}}

\documentstyle{mn}

\input psfig

\title[GB/FIRST High-z quasar Survey]{A Search for High-redshift Quasars Among
GB/FIRST Flat-Spectrum Radio Sources}

\author[I. M. Hook et al.]
{\parbox[]{6.in} 
{I. M. Hook$^1$\thanks{Present address: European Southern
Observatory, Karl Schwarzschild Stra\ss{e} 2, D-85748 Garching b. M\"{u}nchen, Germany}, R. H. Becker$^2$, R. G. McMahon$^3$, R. L. White$^4$} \\ 
      $^1$ U.C. Berkeley Astronomy Dept, Berkeley, CA 94720, U.S.A. \\	 
      $^2$ University of California at Davis, Davis, CA 95616, U.S.A.\\
      $^3$ Institute of Astronomy, Madingley Road, Cambridge CB3 0HA, U.K. \\
      $^4$ Space Telescope Science Institute, 3700 San Martin Drive,
      Baltimore, MD 21218, U.S.A.\\
       {\rm email: ihook@eso.org, bob@igpp.llnl.gov,
      rgm@ast.cam.ac.uk, rlw@stsci.edu}
}

\begin{document}
\maketitle

\begin{abstract} We present the method and first results of a survey
for high-redshift ($z>3$) radio-loud quasars, which is based on
optical identifications of 2902 flat-spectrum radio sources with $\rm
S_{5GHz}\ge 25$mJy. The radio sample was defined over a 1600 sq degree
region using the 5GHz Green Bank survey and the 1.4GHz VLA FIRST
survey. 560 sources were identified to a limit of E=19.5 on APM scans
of POSS-I plates and 337 of these optical counterparts are unresolved.
From these a complete sample of 73 sources for spectroscopic follow up
was defined based on criteria of red ($\rm O-E
\ge 1.2$) optical colour.  We have obtained spectra for 36 of these
and an additional 14 had redshifts in the literature, thus 70\% of the
spectroscopic sample is completed.  Six objects in the sample were
found to be radio-loud quasars with $z>3$ of which two were previously
known.  The efficiency of the spectroscopic phase of the survey is
therefore about 1 in 9, whereas without the colour selection criterion
the efficiency would have been 1 in $\sim 40$.  The six $z>3$ quasars
were found in an effective area of 1100 square degrees, implying a
surface density of one flat-spectrum $z>3$ radio-loud quasar per 190
square degrees to limits of E=19.5 and $\rm S_{5GHz}\ge 25mJy$.  This
survey also produced the first known radio-loud BAL quasar, 1556+3517
with $z=1.48$, which has been reported in an earlier paper (Becker et
al 1997). This object has a redder optical colour ($\rm O-E=2.56$)
than all the $z>3$ quasars found in this survey to date. In addition
we have obtained spectra of 22 GB/FIRST sources which are not part of
the complete sample.  We give positions, E (red) magnitudes, $\rm O-E$
colours, radio fluxes, radio spectral index and redshifts where
possible for objects for which we have obtained spectra. We give
spectra and finding charts for the $z>3$ quasars.

\end{abstract} 
\maketitle

\section{Introduction}

We have begun a search for high-redshift quasars selected using the 
VLA FIRST survey (Becker, White \& Helfand 1995). The aim of this work
is to extend recent searches for radio-loud high-redshift quasars (e.g.
Hook et al. 1995, 1996; Hook \& McMahon 1998; Shaver et al. 1996)
fainter in radio flux density by a factor of $\sim 10$, and, ultimately, to use
the new sample to constrain the faint end of the radio quasar luminosity
function at high redshift. In addition, the sample can be used to
provide an unbiased sample of damped absorption systems that can be
used to study galaxy evolution at high redshift.

Whilst radio-loud quasars are only a small subset of the quasar
population, the construction of radio-loud quasar samples is less prone to
selection effects than are optical samples, since radio emission is
unaffected by either intrinsic or extrinsic absorption due to
dust. Fall \& Pei (1993) discuss how dust within intervening galaxies
may affect the observed evolution in optically selected samples of
quasars.  

Our survey technique involves optical identification of a large number
of radio sources, followed by selection of the optically red,
unresolved objects. The technique is similar to that used by Hook et
al. (1995,1996) to find $\sim 30$ $z>3$ quasars at higher radio flux
densities, and complements the FIRST Bright Quasar sample (FBQS,
Gregg et al. 1996).

The radio data are described in section 2 of this paper and the
optical identification of the radio sample and selection of the
spectroscopic sample are described in section 3. The spectroscopic
observations are described in section 4, and the results are presented
in section 5.  Finally, tables of redshifts, optical and radio data
are given for the objects for which we have obtained spectra, along
with the spectra and finding charts for the $z>3$ quasars.

\section{Definition of the Radio Sample}

This study uses as its starting point the 5GHz 87GB catalog of Gregory
\& Condon (1991). By defining the sample at high frequency,
flat-spectrum core-dominated objects, which are usually identified
with quasars, are preferentially selected.  Spectral index information
and the accurate positions (better than $\pm 1.0''$) needed for
optical identification are provided by the $\rm S_{1.4GHz} \ge 1mJy$
VLA `FIRST' survey (Becker, White \& Helfand, 1995). The sample covers
an area of 0.49sr (1600sq deg), limited by the extent of the FIRST
survey which at the commencement of this work covered the region $\rm 7^h
30^m < \alpha (J2000) < 17^h30^m$, $\rm 28\dgr 30' <\delta (J2000) <
41\dgr 30'$.

Objects with $\rm S > 25 mJy$ from the 5GHz 87GB catalog were matched to
the FIRST catalog with a matching radius of 1 arcmin.  In cases where
more than one FIRST object matched the 5GHz source (about 50\% of the
time), the FIRST source with the strongest flux was selected (i.e the
source with the steepest spectrum). The reason for this is that if
that source then passes the flat spectrum selection criterion then
there is certain to be a flat spectrum source associated with the 5GHz
source.

Matched sources with radio spectral index flatter than $\rm
\alpha^{5GHz}_{1.4GHz} =-0.5$ (where $\rm S\propto \nu^{\alpha}$) were
then selected.  The exact spectral index criterion chosen was somewhat
arbitrary but this has proved successful in the past at separating the
quasar and radio galaxy populations.  The `total' (rather than `peak')
fluxes from the 1.4GHz FIRST catalogue were used to compute the
spectral index. A total of 2902 sources ($\rm \sim2 deg^{-2}$) meet
these criteria.

\section{Optical identification and selection of the Spectroscopic sample}
Optical identification of the radio sources was carried out using APM
(Automated Plate Measurement Facility at Cambridge, U.K.) scans of
POSS-I E (red) and O (blue) plates. The identification procedure and
selection of candidates is similar to that in Hook et al. (1996), and
more details can be found there along with a description of the
APM POSS-I catalog.

Identifications were made based on positional coincidence of the FIRST
radio position with an optical counterpart on the POSS-I E plate.  The
sample of candidate high-redshift quasars was then selected using the
APM image classification, $N\sigma_c$, measured from the E plate, and
the O$-$E colour. The basis for the colour selection method is that
the $\rm O-E$ colours of quasars with $z>2$ become rapidly redder with
redshift, due to absorption by intervening Ly$\alpha$ (see figure 1 in
Hook et al. 1995). To define a sample with a high level of
completeness at $z>3$ we chose a limit of $\rm O-E\ge 1.2$.  A
magnitude limit of $\rm E\le 19.5$ was imposed (0.5 mag brighter than
the E plate limit) so that spectroscopy with the Lick-3m would be more
feasible. The APM star/galaxy separation is also more reliable away
from the plate limit.

\begin{figure*}
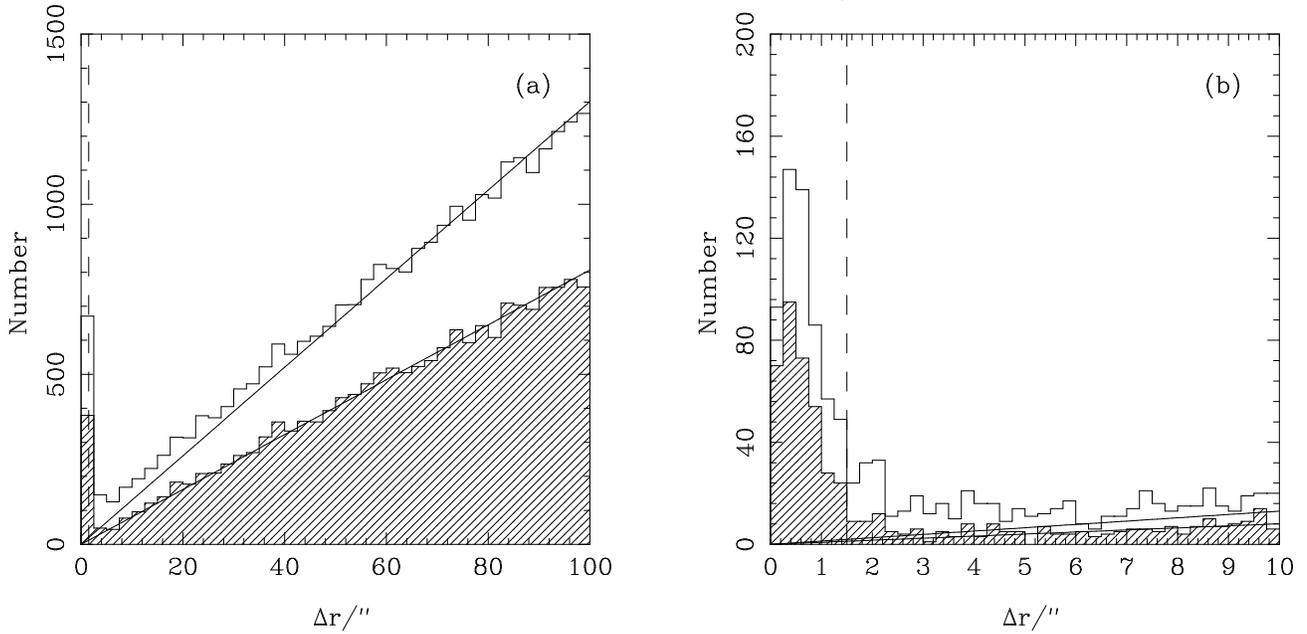

\centerline{
\psfig{figure=fig1a.ps,height=3.2in,bbllx=34pt,bblly=447pt,bburx=272pt,bbury=689pt}\hspace{1.2cm}\psfig{figure=fig1b.ps,height=3.2in,bbllx=34pt,bblly=447pt,bburx=272pt,bbury=689pt}
}
\caption[]{Histograms of positional difference $\Delta r$ between the
optical position measured by the APM and the FIRST radio position for
the flat spectrum sample. (b) shows a blowup of the
region $\Delta r \le 10.0''$. The open histogram shows all
identifications with $\rm E\le19.5$ and the shaded histogram shows
those which are classified as stellar ($N\sigma_c\le 2.0$, see
text). The dashed line shows the limit used to define the
spectroscopic sample, $\Delta r \le 1.5''$. The solid sloping lines
show fits to the histograms in the range $40.0'' < \Delta r < 100.0''$,
which determine the background counts. The upper line corresponds to
1.03 background sources with $\rm E\le19.5$ per $\rm arcmin^2$. The
lower line corresponds to 0.64 stellar sources per $\rm arcmin^2$ to
the same magnitude limit.}
\label{delr}
\end{figure*}

Based on results from the $\rm S\ge 200mJy$ high-redshift
radio-quasar survey of Hook et al. (1996) and on results from from the
FBQS (Gregg et al.  1996), stricter selection criteria were used for
this new survey. The criterion for positional coincidence used by Hook et
al. (1996) was $\Delta r \le 3.0''$ and for the FBQS it was $\Delta r
\le 2.0''$ (Gregg et al. 1996).  Even that was found to be too
generous: 64 out of 69 (93\%) spectroscopically confirmed quasars in
FBQS were found to have $\Delta r \le 1.0''$, and 97\% have $\Delta r
\le 1.1''$ (the outliers were extended radio sources). Since the
current high redshift survey goes 2 magnitudes fainter in E than FBQS,
we adopt a more conservative criterion of $\Delta r \le 1.5''$.
Figure~\ref{delr} shows this limit in relation to the histogram of
$\Delta r$ for identifications with $\rm E \le 19.5$.  The
distribution peaks for the stellar and non-stellar identifications at
$\Delta r \sim 0.5''$. However, whereas beyond $\sim 3''$ the stellar
histogram is indistinguishable from the derived background, the
non-stellar identifications show an excess. There is evidence from
Figure~\ref{delr}a that this may persist out to a radius of $\sim
40''$ (and possibly further, although this would not be apparent in
Fig~\ref{delr}a since the background was determined by fitting the
distribution from $40''$ to $100''$). This may be due to either the
clustering of galaxies or that the peak of the radio emission is not
on the optical counterpart as would be expected for extended asymmetric
radio sources. Since the extended or resolved optical counterparts are
not the subject of this paper, we do not consider this issue any
further.  This issue is discussed further in a forthcoming paper
(McMahon et al, in prep) on the optical identifications of all FIRST
sources using the APM POSS-I catalogue.

There is a small excess of 34 stellar objects above the background
level between radii of $1.5''$ and $3.0''$ (see Fig~\ref{delr}b). If we
assume that these are real identifications that are excluded from
the spectroscopic sample, then the incompleteness due to the $\Delta r
\le 1.5''$ criterion is is $\sim 9\%$. This agrees with the fact that of the 13
$z>3$ quasars found in the survey of Hook et al. (1996) (with limiting
magnitude $\rm E\le 20$), only one would not have satisfied the
criterion of $\Delta r \le 1.5''$. 

The background counts for stellar sources with $\rm E\le 19.5$ was
found to be 0.64 per square arcmin (see Fig.~\ref{delr} and
caption). Thus within a search radius of $1.5''$ we expect 3.6 false
identifications amongst the stellar identifications, i.e. 1.1\%.  The
corresponding background for red ($\rm O-E \ge 1.2$) stellar sources
is 0.54 per square arcmin, and we therefore expect 3.1 false red
stellar identifications. In section 5 this number is compared to the
number of stars found in our sample.

We also now adopt a stricter criterion for stellar image
classification.  In Hook et al. (1996) a criterion of $|N\sigma_c| \le
3.0$ was used (where $N\sigma_c$ can be considered as the number of
standard deviations a particular object is from having a stellar
profile; see Hook et al. (1996) for a more detailed
explanation). However only one of the 13 $z>3$ quasars found in that
survey had $|N\sigma_c| > 2.0$, an object with E=19.93, fainter than
the limit of the current survey. Thus we now adopt a criterion of
$|N\sigma_c| \le 2.0$. Whilst using these stricter criteria on $\Delta
r$ and $N\sigma_c$ has little effect on the completeness of the final
high-redshift quasar sample compared to that in Hook et al. (1996), the
number of candidates requiring spectroscopy is reduced by $\sim 40\%$.

In summary the selection criteria are (i) $\rm S_{5GHz} \ge 25mJy$, (ii)
flat radio spectral index $\rm \alpha^{5GHz}_{1.4GHz} \ge -0.5 $, (iii)
positional coincidence, $\Delta r\le 1.5''$, (iv) red optical colour
$\rm O-E\ge 1.2$, (v) unresolved image on the E plate, $|N\sigma_c|\le
2.0$, (vi) $\rm E \le 19.5$ mag.  The optical counterpart was not
required to be detected on the O plate.

There is little overlap between this sample and the FBQS. The FBQS does
not have a radio spectral index cut and reaches $\rm S_{1.4GHz} \sim
1mJy$. FBQS now has a blue optical colour criterion of $\rm O-E < 2.0$
and has a significantly brighter optical limit of E=17.5. Hence the
two quasar samples are largely complementary.

\begin{figure}
\centerline{
\psfig{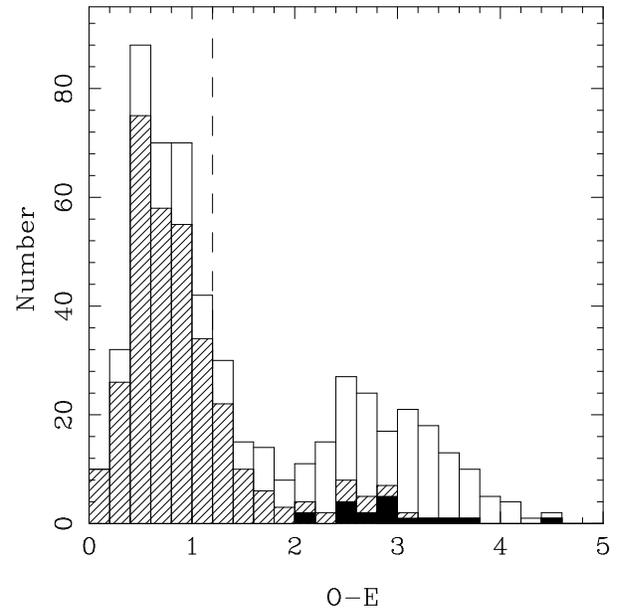}}
\caption[]{Colour histogram for APM identifications of flat spectrum
GB/FIRST sources. The open histogram shows all identifications (both
stars and galaxies) within $1.5''$ and with $\rm E\le19.5$. The hatched
histogram shows the stellar sources only (337) and the solid histogram
shows the stellar sources which were not detected on the O plate and
for which the $\rm O-E$ values are limits. The dashed line shows the
colour selection
criterion adopted ($\rm O-E\ge 1.2$), which eliminates the blue
stellar identifications (i.e. low-redshift quasars) from the
spectroscopic sample.}
\label{oe_hist}
\end{figure}

\begin{figure}
\centerline{
\psfig{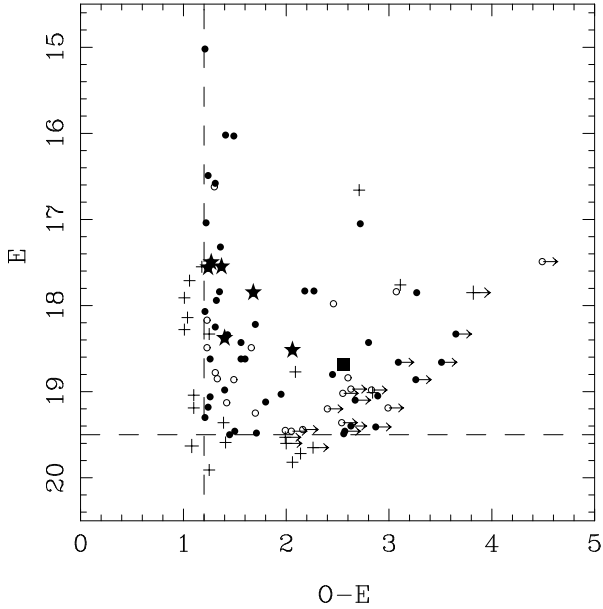}
}
\caption[]{Colour-magnitude diagram for APM
identifications of flat-spectrum GB/FIRST sources. The circles
represent objects in the spectroscopic sample: those with
spectroscopic data are shown as filled circles while those with no
spectra as yet are shown as open circles.  Arrows show limits on $\rm
O-E$ colour for objects that were not detected on the O plate. The
$z>3$ quasars are shown by stars and the radio-loud BAL quasar
1556+3517 is shown by the square. The dashed lines show the E and $\rm
O-E$ limits of the spectroscopic sample. The extra objects observed
which do not satisfy all the selection criteria (see section 5) are
shown as crosses.}
\label{rqcolmag}
\end{figure}

Of the 2902 flat spectrum sources (as defined in section 2) in the
GB/FIRST survey area, 560 have optical identifications brighter than
E=19.5 within $1.5''$.  The histogram of their $\rm O-E$ colours, shown
in Figure~\ref{oe_hist}, has a bimodal distribution. The bluer peak
is dominated by stellar identifications (the vast majority of which
are low-redshift quasars) and the redder peak is dominated by
galaxies. By imposing a red colour selection criterion on
the sample of stellar identifications, the majority of low redshift
quasars are eliminated. Of the 560 identifications, 337 are classified
as stellar and 73 of these satisfy the colour criterion above ($\rm
O-E \ge 1.2$). This sample of 73 objects will be referred to as the
`spectroscopic sample' and they are shown in the colour-magnitude
diagram in Figure~\ref{rqcolmag}.

\begin{figure}
\centerline{\psfig{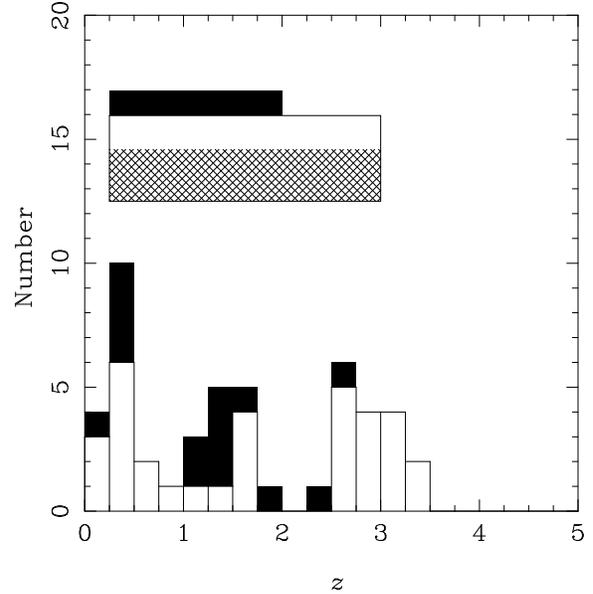}}
\caption[]{Histogram of redshifts obtained.
The white histogram represents the spectroscopic sample, including
previously-known objects and objects with uncertain redshifts, a total
of 33 objects (the two stars are not included here). Objects in the
spectroscopic sample with spectra from which redshifts could not be
measured (15 objects - it is unlikely that any of these have $z>3$;
see text for explanation) are represented by the white box and objects
for which spectra have not been obtained (23) are represented by the
hatched box. Similarly the black histogram and rectangle show the the
extra objects observed which satisfy a less stringent set of selection
criteria described in section 5 (15 of these have definite redshifts
and 7 have `featureless' spectra).}
\label{zhist}
\end{figure}

Fourteen of the spectroscopic sample had published spectroscopic data,
leaving 59 objects requiring spectroscopy.  The previously-known
objects were not re-observed and their properties are summarised in
Table~1. Two of the previously-known objects are radio-selected $z>3$
quasars (0933+2845, Gregg et al. 1996 and 1340+3754, Hook et
al. 1995). 

\begin{table*}
\begin{center}
\caption{Optical and radio data for objects in the
spectroscopic sample which had spectroscopic data in the literature.}
\begin{tabular}{r@{\hspace{0.2cm}}     r@{\hspace{0.2cm}}rr@{\hspace{0.2cm}}r@     {\hspace{0.2cm}}rlrrrrrrrrrl}
\hline
\multicolumn{6}{c}{Optical Position}&       \multicolumn{1}{c}{$z$} &
$N\sigma_c$    &\multicolumn{1}{c}{E}       & O$-$E &$\Delta r$
&\multicolumn{1}{c}{$\rm S_{\scriptscriptstyle GB}$}       &
$\rm S_{\scriptscriptstyle B91}$  &
$\rm S_{\scriptscriptstyle FIRST}$& $\rm S_{\scriptscriptstyle NVSS}$ &\multicolumn{1}{c}
{$\rm \alpha^{5GHz}_{1.4GHz}$} &Reference/ \cr
\multicolumn{2}{c}{$\alpha$}&   \multicolumn{2}{c}{J2000}&
\multicolumn{2}{c}{$\delta$}& &&mag &mag& \multicolumn{1}{c}{$''$}
&mJy&mJy &mJy&mJy& & Classification\cr\hline
  09&11&47.71& +33&49&17.6&   $-$&$   1.75$&  18.62&$    1.56$&  0.78&
250&     207&   391.2&    380.8\phantom{$^a$}&$   -0.35$&H96 / featureless\cr
  09&30&55.28& +35&03&36.9&   $-$&$   0.12$&  19.18&$    1.24$&  0.85&
383&     381&   518.8&    484.5\phantom{$^a$}&$   -0.24$&H96 / featureless  \cr
  09&33&37.31& +28&45&32.3&  3.42&$   0.27$&  17.50&$    1.27$&  0.16&
66&      63&   120.9&    117.0\phantom{$^a$}&$   -0.48$&G96 FBQS / QSO  \cr
  10&50&58.15& +34&30&10.8&  2.52&$   1.43$&  18.98&$    1.40$&  0.30&
298&     291&   547.3&    554.0\phantom{$^a$}&$   -0.48$&W84 / QSO  \cr
  11&49&00.29& +29&58&41.0&  0.15&$   1.15$&  17.05&$    2.72$&  0.34&
41&      39&    53.2&     52.5\phantom{$^a$}&$   -0.20$&G96 FBQS / Gal\cr
  12&35&05.79& +36&21&18.7&  1.60&$  -0.20$&  19.06&$    1.26$&  0.47&
250&     251&   175.7&    154.5\phantom{$^a$}&$    0.28$&H96 / QSO   \cr
  13&24&12.09& +40&48&12.4&  0.50&$   0.16$&  19.49&$    2.56$&  0.87&
413&     423&   369.4&    350.1\phantom{$^a$}&$    0.09$&V96 / QSO \cr
  13&40&22.94& +37&54&44.4&  3.10&$   0.08$&  17.85&$    1.68$&  0.85&
305&     301&   284.3&    273.9\phantom{$^a$}&$    0.06$&H95 / QSO  \cr
  13&50&35.95& +33&42&17.2&  0.01&$  -0.85$&  11.41&$    1.35$&  0.93&
76&      90&   100.5&103.8$^a$&$   -0.22$&HGC95 / Gal \cr
  14&04&14.63& +28&46&36.5&   $-$&$  -0.84$&  16.03&$    1.49$&  1.03&
38&      36&    12.5&     11.7\phantom{$^a$}&$    0.87$&G96 FBQS / star  \cr
  14&21&25.62& +39&43&29.4&  0.62&$   1.43$&  18.33&$  > 3.65$&  0.79&
67&      64&    43.6& 82.9$^b$&$    0.34$&V90 / QSO               \cr
  14&58&44.88& +37&20&21.6&  0.33&$  -0.67$&  18.22&$    1.70$&  0.68&
591&     591&   271.9&    215.2\phantom{$^a$}&$    0.61$&V96, H96 / Gal          \cr
  15&58&55.21& +33&23&19.3&  1.65&$  -0.13$&  16.49&$    1.24$&  0.83&
88&      86&   145.5&    142.7\phantom{$^a$}&$   -0.39$&F86 / QSO      \cr
  16&33&02.12& +39&24&27.4&  1.02&$   0.25$&  15.02&$    1.21$&  0.34&
40&      38&    54.7&     79.1\phantom{$^a$}&$   -0.25$&CCH90 / QSO    \cr
\hline
\end{tabular}
\end{center}
\label{known}
\raggedright
\noindent Notes on column headings: $\rm S_{GB}$ is the 5GHz flux from
Gregory \& Condon (1991) and $\rm S_{B91}$ is the 5GHz flux from the
Becker et al (1991) catalog.  $\rm S_{NVSS}$ was determined by summing up
the fluxes from sources in the 1.4GHz NRAO-VLA Sky Survey catalog ({\it
NVSS}, Condon et al 1994) within $60''$ of the FIRST position.
$\rm S_{FIRST}$ is the `Total' flux from the 1.4GHz FIRST catalog and
$\rm \alpha^{5GHz}_{1.4GHz}$ is the spectral index between $\rm S_{FIRST}$ and
$\rm S_{GB}$.\newline
$^a$ NVSS flux is the sum of two components. The central component has
S=97.3mJy and a second component $20''$ away has S=6.5mJy.\newline
$^b$ NVSS flux is the sum of two components: The closest source has 
S=62.3mJy and a second, probably unrelated source $42''$ away has S=20.6mJy.\newline
\noindent References:
F86=Foltz et
al (1986), G96 = Gregg et al (1996), H95 = Hook et al (1995), H96 = Hook et al (1996), HGC95=
Huchra, Geller \& Corwin (1995), V90=Vigotti et al (1990), V96 =
Vermeulen et al (1996), W84=Wampler et al 1984, CCH90=Crampton, Cowley
\& Hartwick (1990) \newline
\end{table*}

\section{Optical Spectroscopy} The aim of the spectroscopic phase of
this project is to obtain classification (high-redshift quasar or not)
for the complete spectroscopic sample defined above.  This does not
require redshifts to be obtained for all the objects, for example many
`featureless' spectra could be rejected as possible $z>3$ quasars by
the lack of Ly$\alpha$ forest absorption.  The strategy was therefore
to take short exposures at low dispersion to obtain spectra of
sufficient quality for the nature of the object and, where possible,
the redshift to be determined.

Most of the spectroscopy was carried out in 1995 April and 1996 May
using the KAST double-spectrograph at the Cassegrain focus of the 3m
Shane Telescope at Lick observatory.  Typical exposure times were
600-900s.
For the May 1996 run, a grating giving 4.6\AA/pix dispersion was used
on the red side, and on the blue side the selected grism gave a
dispersion of 2.54\AA/pix.  A dichroic at $\sim$5500\AA\ was used.
The set-up was such that the blue side of the spectrum covered
$2900-6000$\AA\ and the red side covered $5080-10610$\AA.
For the Apr 1995 run a grating giving 2.32\AA/pix dispersion was used
on the red side, and on the blue side the selected grism gave a
dispersion of 1.85\AA/pix.  A dichroic at $\sim$5500\AA\ was used
giving a total wavelength coverage of 3430-8150\AA.  For both runs the
detectors on both the red and blue sides were $1200\times 400$
UV-flooded Reticon CCDs with 27 micron pixels, corresponding to about
0.8 arcsec/pixel. The CCD chips were windowed in the spatial direction
to reduce the readout time.

Seven objects from the spectroscopic sample were observed at the Keck
telescope with the Low Resolution Imaging Spectrograph (LRIS). These
observations were carried out as a backup program during a run to
obtain spectra of distant supernovae.  A 300 l/mm grating was used on
LRIS and covered the wavelength range $4900-9900$\AA\ with resolution
of 2.5\AA/pix.  In addition, one object was observed at 4.2m WHT, La
Palma in 1993 April and another by R. Weymann at Palomar 5m in 1991
December as part of other programs.

For all observations the width of the slit was adjusted to be
compatible with the seeing at the time of observation ($1-2''$) and
the data were taken with a long slit at the parallactic angle. Some of
the objects were too faint to be easily visible on the acquisition TV
at the Lick 3m so accurate offsetting from nearby stars was used to
position the target object in the slit.  Spectrophotometric standards
were observed to calibrate the spectra.

 The reduction of data was carried out using standard software from
the IRAF\footnote{IRAF is distributed by the National Optical
Astronomy Observatories, which is operated by the Association of
Universities for Research in Astronomy, Inc. (AURA) under cooperative
agreement with the National Science Foundation.} package.

\section{Results and Discussion}

Of the 59 objects requiring spectroscopy we now present new
spectroscopic data for 36. The properties of objects in the
spectroscopic sample for which we obtained spectra are given in
Table~2. When combined with the 14 objects which have
spectroscopic data in the literature, 50 ($70\%$) of the complete
sample defined in Section 3 now have spectroscopic data. The remaining
23 objects have yet to be observed.

\begin{table*}
\begin{center}
\caption{Optical and radio data for objects from the complete sample
observed spectroscopically. These objects satisfy the selection
criteria (i) $\rm S_{5GHz, GB} \ge 25mJy$, (ii) flat radio spectral index
$\rm \alpha^{5GHz}_{1.4GHz} \ge -0.5 $, (iii) positional coincidence, $\Delta r\le
1.5''$, (iv) red optical colour $\rm O-E\ge 1.2$, (v) unresolved image
on the E plate, $|N\sigma_c|\le 2.0$, (vi) $\rm E \le 19.5$ mag.  }
\begin{tabular}{r@{\hspace{0.2cm}}     r@{\hspace{0.2cm}}rr@{\hspace{0.2cm}}r@     {\hspace{0.2cm}}rlrrrrrrrrrl}
\hline
\multicolumn{6}{c}{Optical Position}&       \multicolumn{1}{c}{$z$} &
$N\sigma_c$    &\multicolumn{1}{c}{E}       & O$-$E &$\Delta r$
&\multicolumn{1}{c}{$\rm S_{\scriptscriptstyle GB}$}       &  $\rm
S_{\scriptscriptstyle B91}$  & $\rm S_{\scriptscriptstyle FIRST}$&
$\rm S_{\scriptscriptstyle NVSS}$ &\multicolumn{1}{c}
{$\rm \alpha^{5GHz}_{1.4GHz}$} &Tel/Date/Class \cr
\multicolumn{2}{c}{$\alpha$}&   \multicolumn{2}{c}{J2000}&   \multicolumn{2}{c}{$\delta$}& &&mag &mag&     $''$ &mJy&mJy &mJy&mJy&\cr\hline
  08&47&15.16& +38&31&10.2&  3.18&$  -1.01$&  17.55&$    1.37$&  0.31&    124&        122&   185.8&    201.1\phantom{$^a$}&$   -0.32$&K 03/97 QSO        \cr
  08&52&44.76& +34&35&40.8&  1.65&$  -1.31$&  18.43&$    1.56$&  0.53&     70&         66&    69.6& 69.1$^a$&$    0.01$&K 03/97 QSO        \cr
  08&57&41.88& +38&16&35.6&  0.43&$   1.15$&  19.40&$  > 2.63$&  0.47&     49&         47&    33.8& 70.9$^b$&$    0.29$&K 03/97 Gal (abs.)  \cr
  08&59&17.31& +34&09&09.3&  0.55&$   1.81$&  19.46&$  > 2.57$&  0.65&     30&           &    55.6&     53.8\phantom{$^a$}&$   -0.48$&K 03/97 Gal (abs.)  \cr
  09&02&39.91& +39&57&26.7&  3.08&$   0.83$&  17.56&$    1.24$&  0.14&     31&         29&    32.4&         \phantom{$^a$}&$   -0.03$&K 03/97 QSO       \cr
  10&23&24.04& +28&56&50.8&   $-$&$   1.63$&  19.12&$    1.80$&  0.24&     89&         86&   127.4&    139.7\phantom{$^a$}&$   -0.28$&W 04/93 featureless        \cr
  10&28&46.95& +41&26&56.6& 2.82*&$   1.11$&  19.46&$    1.50$&  0.19&     53&         51&    62.3&     74.4\phantom{$^a$}&$   -0.13$&L 05/96 QSO ?  \cr
  10&42&06.79& +29&38&26.8&  2.55&$  -0.09$&  19.30&$    1.21$&  0.73&     56&         54&    70.0&     71.9\phantom{$^a$}&$   -0.18$&L 04/95 QSO         \cr
  10&42&32.27& +36&15&20.7&   $-$&$   1.81$&  18.62&$    1.60$&  0.09&     92&         89&    60.1&     68.1\phantom{$^a$}&$    0.33$&L 05/96 featureless ?  \cr
  10&44&06.39& +29&59&01.3&  2.99&$   0.61$&  18.62&$    1.26$&  0.73&     80&         79&    37.4&     40.9\phantom{$^a$}&$    0.60$&P 12/91 QSO        \cr
  10&55&38.67& +30&52&50.6&   $-$&$   1.69$&  17.83&$    2.27$&  0.65&     32&         31&     7.8&     14.7\phantom{$^a$}&$    1.11$&L 04/95 not QSO       \cr
  11&21&11.12& +37&02&47.0&  1.62&$  -1.09$&  17.04&$    1.22$&  0.16&     72&         69&   100.8&    105.7\phantom{$^a$}&$   -0.26$&L 05/96 QSO   \cr
  11&29&14.28& +37&03&17.5&   $-$&$  -0.99$&  18.34&$    1.43$&  0.52&     95&         91&    70.1&     60.9\phantom{$^a$}&$    0.24$&L 05/96 featureless    \cr
  11&51&16.95& +40&08&21.6&  2.74&$  -0.88$&  19.41&$  > 2.87$&  0.52&    111&        108&    84.0&     81.7\phantom{$^a$}&$    0.22$&L 05/96 QSO            \cr
  12&34&44.34& +36&37&22.2&   $-$&$  -1.07$&  19.10&$  > 2.67$&  0.27&     45&         43&    62.8&     64.3\phantom{$^a$}&$   -0.26$&L 05/96 Gal (abs?)    \cr
  12&36&23.02& +39&00&00.7&   $-$&$   0.01$&  17.32&$    1.36$&  0.32&     47&         57&    37.0&     37.1\phantom{$^a$}&$    0.19$&L 05/96 featureless  \cr
  12&58&32.18& +29&09&03.0&  3.47&$   0.32$&  18.52&$    2.06$&  0.48&     32&         32&    25.0&     25.1\phantom{$^a$}&$    0.19$&L 04/95 QSO   \cr
  13&01&23.40& +28&42&53.8&  0.42&$   1.40$&  18.86&$  > 3.26$&  0.47&     31&           &    13.1&     12.1\phantom{$^a$}&$    0.68$&L 04/95  Gal (em.)\cr
  13&24&04.90& +34&17&38.5&   $-$&$   1.64$&  18.80&$    2.45$&  0.35&     85&         83&    54.8&     62.2\phantom{$^a$}&$    0.34$&L 05/96 featureless ?  \cr
  13&42&52.97& +40&32&02.0&  0.91&$   0.99$&  19.50&$    1.45$&  0.58&    113&         99&   137.2&    152.2\phantom{$^a$}&$   -0.15$&L 05/96 QSO            \cr
  13&48&48.13& +29&39&17.6&  0.24&$   2.00$&  18.43&$    2.80$&  1.19&
216&        213&   327.5&    695.6\phantom{$^a$}&$   -0.33$&L 04/95
Gal \cr 
  13&58&19.85& +32&19&27.8&  2.85&$  -0.27$&  18.66&$  > 3.09$&  1.32&     36&         36&    20.6&     18.8\phantom{$^a$}&$    0.44$&L 05/96 QSO \cr
  14&01&38.82& +36&11&21.6&   $-$&$   0.60$&  19.03&$    1.95$&  1.29&     61&         59&    83.3&     73.0\phantom{$^a$}&$   -0.24$&L 05/96 featureless    \cr
  14&13&14.15& +32&00&56.6&   $-$&$   1.80$&  19.05&$    2.89$&  1.19&     31&           &    44.0&     49.0\phantom{$^a$}&$   -0.28$&L 05/96 featureless  \cr
  14&36&42.84& +40&22&05.8&  2.57&$  -0.93$&  18.07&$    1.21$&  0.19&     33&         32&    23.0&     22.3\phantom{$^a$}&$    0.28$&L 05/96 QSO            \cr
  14&53&18.56& +35&05&39.4&   $-$&$  -0.13$&  19.48&$    1.71$&  0.51&    127&        123&   215.8&    217.3\phantom{$^a$}&$   -0.42$&L 05/96 featureless    \cr
  15&23&14.87& +38&14&01.7&  3.15&$  -1.09$&  18.38&$    1.40$&  0.40&     27&           &    47.7&     52.6\phantom{$^a$}&$   -0.45$&L 05/96 QSO            \cr
  15&29&42.17& +35&08&50.7&  0.29&$   0.85$&  17.83&$    2.18$&  0.94&     50&         48&    90.0&    105.6\phantom{$^a$}&$   -0.46$&L 05/96 QSO  \cr
  15&42&46.77& +33&46&02.4&  0.28&$   1.99$&  17.85&$    3.27$&  0.93&     69&         66&    36.8&     39.8\phantom{$^a$}&$    0.49$&L 05/96 Gal   \cr
  15&56&33.79& +35&17&56.9&  1.48&$  -1.71$&  18.68&$    2.56$&  0.71&     27&           &    31.8&     28.1\phantom{$^a$}&$   -0.13$&L 05/96 BAL QSO   \cr
  16&05&23.73& +30&38&37.0&  2.67&$   0.27$&  18.66&$  > 3.51$&  0.44&     76&         59&    52.6&     50.0\phantom{$^a$}&$    0.29$&L 04/95 QSO \cr
  16&20&04.77& +35&15&54.5&  2.95&$   0.91$&  17.94&$    1.32$&  1.01&     40&         38&    50.1&     57.9\phantom{$^a$}&$   -0.18$&L 05/96 QSO            \cr
  16&55&42.82& +32&44&20.0&   $-$&$  -1.37$&  17.84&$    1.35$&  0.30&     69&         69&    31.8&     38.7\phantom{$^a$}&$    0.61$&L 05/96 featureless ?  \cr
  17&00&08.72& +29&19&03.1&   $-$&$  -2.00$&  16.02&$    1.41$&  1.08&     39&         48&     3.0&      5.1\phantom{$^a$}&$    2.01$&L 05/96 star   \cr
  17&01&24.70& +39&54&37.3&   $-$&$   0.64$&  16.58&$    1.31$&  0.82&    159&        157&   256.8&    190.8\phantom{$^a$}&$   -0.38$&L 05/96 featureless \cr
  17&06&50.43& +30&04&12.1&   $-$&$  -0.03$&  18.25&$    1.31$&  0.32&     38&         36&    27.9&     29.5\phantom{$^a$}&$    0.24$&L 05/96 featureless  \cr
\hline
\end{tabular}
\end{center}
\label{spectab}
\raggedright
\noindent{\bf Notes}\newline
* uncertain redshift. \newline
Telescope code: L=Lick 3m,
K=Keck-II, W=WHT, P=Palomar 5m (Weymann, private communication).\newline
$^a$ NVSS flux contains contributions from two components: a central
component with flux density S=60.7mJy and another $14''$ away with S=8.4mJy.\newline
$^b$ As above but central component has S=54.5mJy and other component
$36''$ away has S=16.2mJy.
\end{table*}

Four new quasars with $z>3$ have been found in the spectroscopic
sample to date. Since 36 objects were observed, the efficiency of the
survey for finding $z>3$ quasars is approximately 1 in 9, similar to
that of Hook et al 1996 (although note that the selection criteria are
now stricter).  If we had not used the colour selection criterion, the
efficiency would be lower by a factor of $\sim 4.5$. The $z>3$ quasars
are marked on the colour-magnitude diagram (Figure~\ref{rqcolmag})
with stars.  A further 8 new quasars in the range $2<z<3$ were found
(but note that the colour selection criterion means the sample is
substantially incomplete for $z<3$).

The remaining observed objects were found to be low-redshift ($z<2$)
quasars, emission- or absorption-line galaxies and objects with
featureless spectra which could be ruled out as possible high-redshift
quasars.  Two objects were found to be stars and are probably chance
coincidences with the radio position (both have $\Delta r > 1.0''$).
This agrees well with the number of false identifications expected
(see section 3). We expect 3.1 false, red, stellar identifications in
the full spectroscopic sample, thus after observing 70\% of these
objects we expect to have found 2.1 stars.

The first radio-loud BAL quasar, 1556+3517 with $z=1.48$, was
discovered in this survey and was reported in an earlier paper (Becker
et al. 1997). Note that this object is redder ($\rm O-E=2.56$) than
any of the $z>3$ quasars in the sample so far.

Of the 50 objects with spectroscopic data, 35 have a definite redshift
based on emission or absorption features (including the two
stars). Their redshift distribution is shown in Figure~4.  The 15
`featureless' objects are represented by the open box, while objects
with no spectral information are represented by the hatched box.

In addition we have observed 22 objects that do not meet one or more
of the selection criteria of the spectroscopic sample, but which
satisfy a less stringent set of selection criteria: $\Delta r\le
3.0''$; $\rm O-E\ge 1.0$; $|N\sigma_c|\le 3.0$; $\rm E \le 20$ mag.
None of them was found to have $z>3$. The results for these objects
are summarised in Table~3. They are plotted as crosses on the
colour-magnitude diagram in Figure~\ref{rqcolmag} and represented by
the solid shading in the redshift histogram in Figure~4.

\begin{table*}
\begin{center}
\caption{Optical and radio data for objects observed spectroscopically
that were not part of the complete spectroscopic sample. These objects
satisfy a less stringent set of selection criteria: $\Delta r\le
3.0''$; $\rm O-E\ge 1.0$; $|N\sigma_c|\le 3.0$; $\rm E \le 20$ mag.}
\begin{tabular}{r@{\hspace{0.2cm}}     r@{\hspace{0.2cm}}rr@{\hspace{0.2cm}}r@     {\hspace{0.2cm}}rlrrrrrrrrrl}
\hline
\multicolumn{6}{c}{Optical Position}&       \multicolumn{1}{c}{$z$} &
$N\sigma_c$    &\multicolumn{1}{c}{E}       & O$-$E &$\Delta r$
&\multicolumn{1}{c}{$\rm S_{\scriptscriptstyle GB}$}       &  $\rm
S_{\scriptscriptstyle B91}$  & $\rm S_{\scriptscriptstyle FIRST}$&
$\rm S_{\scriptscriptstyle NVSS}$ &\multicolumn{1}{c}
{$\rm \alpha^{5GHz}_{1.4GHz}$} &Tel/Date/Class \cr
\multicolumn{2}{c}{$\alpha$}&   \multicolumn{2}{c}{J2000}&   \multicolumn{2}{c}{$\delta$}& &&mag &mag&     $''$ &mJy&mJy &mJy&mJy&\cr\hline
  08&35&19.76& +29&09&21.2&  1.27&$   1.37$&  19.72&$    2.14$&  0.55&     36&      45&    41.3&     38.9\phantom{$^a$}&$   -0.11$&L 04/95 QSO   \cr
  09&00&30.44& +33&30&09.7&   $-$&$   1.09$&  19.91&$    1.25$&  0.86&     81&      77&    72.4&     82.5\phantom{$^a$}&$    0.09$&K 03/97 featureless    \cr
  09&06&02.49& +41&16&27.6&  0.41&$   3.00$&  17.85&$  > 3.82$&  2.75&    279&     276&   771.1&    848.3\phantom{$^a$}&$   -0.80$&K 03/97 QSO            \cr
  09&08&53.08& +30&47&38.3&  1.91&$   0.65$&  19.63&$    1.08$&  0.27&     35&      32&    34.2&     40.1\phantom{$^a$}&$    0.02$&L 04/95  QSO        \cr
  09&24&38.22& +30&28&36.6&  0.27&$   2.95$&  17.76&$    3.11$&  0.70&     27&        &     4.0&      8.4\phantom{$^a$}&$    1.50$&L 04/95  Gal\cr
  09&27&39.78& +30&34&17.3&  1.20&$  -0.24$&  19.59&$    1.41$&  0.87&     57&      56&    46.1&     51.5\phantom{$^a$}&$    0.17$&L 04/95  QSO        \cr
  11&05&50.05& +30&01&46.2&   $-$&$   2.07$&  19.36&$    1.39$&  0.79&     53&      50&    64.2&         \phantom{$^a$}&$   -0.15$&L 04/95 featureless    \cr
  11&24&29.65& +28&31&25.3&  1.35&$   1.08$&  19.04&$    1.10$&  0.46&     31&      38&    22.3&     24.0\phantom{$^a$}&$    0.26$&L 04/95 QSO         \cr
  11&27&29.21& +37&56&12.4&   $-$&$  -0.09$&  19.60&$  > 2.00$&  1.13&     70&      67&    66.4&     78.3\phantom{$^a$}&$    0.04$&L 05/96 featureless ?  \cr
  12&11&34.02& +28&47&33.1&   $-$&$  -0.81$&  19.65&$  > 2.26$&  0.17&     35&      33&    32.9&     32.0\phantom{$^a$}&$    0.05$&L 04/95 ?    \cr
  12&24&09.80& +28&35&10.3&  0.40&$   0.89$&  19.82&$    2.06$&  0.84&     77&      74&    18.6& 297.8$^a$&$    1.12$&L 04/95 Gal (em.)  \cr
  12&39&14.99& +40&49&55.2&  1.31&$  -0.37$&  17.55&$    1.18$&  1.02&     78&      74&    65.7&     70.3\phantom{$^a$}&$    0.13$&L 05/96 QSO            \cr
  13&01&56.57& +29&04&55.7&  2.29&$  -0.57$&  19.19&$    1.10$&  0.44&     50&      48&    97.5&    103.1\phantom{$^a$}&$   -0.52$&L 04/95 QSO  \cr
  13&24&57.44& +32&51&46.2&  1.74&$   0.11$&  17.91&$    1.01$&  1.21&     92&      92&    57.6&     56.7\phantom{$^a$}&$    0.37$&L 05/96 QSO \cr
  13&26&13.76& +28&51&43.6&  1.29&$   1.25$&  18.14&$    1.04$&  0.69&     31&      46&    20.9&     36.8\phantom{$^a$}&$    0.31$&L 04/95 QSO         \cr
  14&08&35.27& +36&17&59.7&   $-$&$  -1.09$&  18.28&$    1.01$&  0.31&     82&      79&   128.3&    133.3\phantom{$^a$}&$   -0.35$&L 05/96 featureless    \cr
  14&13&13.65& +41&16&37.8&  2.61&$   1.41$&  17.71&$    1.06$&  0.34&     35&      33&    26.7&     30.3\phantom{$^a$}&$    0.21$&L 05/96 QSO            \cr
  14&36&27.24& +41&29&31.6& 0.17*&$   1.01$&  18.77&$    2.09$&  1.63&     41&      38&    45.6&     53.6\phantom{$^a$}&$   -0.08$&L 05/96 Gal       \cr
  14&45&27.10& +30&51&28.6&  0.42&$   2.51$&  19.01&$    2.84$&  0.57&     35&     107&     6.3&     12.7\phantom{$^a$}&$    1.35$&L 04/95 Gal (em.)\cr
  15&36&27.97& +37&42&05.9&   $-$&$  -0.41$&  19.53&$  > 1.99$&  1.00&     41&      49&    47.0&         \phantom{$^a$}&$   -0.11$&L 05/95 featureless    \cr
  17&03&22.63& +38&59&29.0&  1.08&$   1.73$&  18.33&$    1.25$&  1.56&     52&      49&    97.4&    104.9\phantom{$^a$}&$   -0.49$&L 05/96 QSO            \cr
  17&07&35.97& +35&39&50.0&   $-$&$   2.13$&  16.66&$    2.71$&  0.51&     53&      51&   100.1&         \phantom{$^a$}&$   -0.50$&L 05/96 featureless    \cr
\hline
\end{tabular}
\end{center}
\label{extratab}
\raggedright
\noindent * uncertain redshift. \newline
\noindent $^a$ The source has complex structure in the FIRST map. The
FIRST flux given is the flux of the central component only. The NVSS
flux is the sum of 4 components within $60''$, the most central of which
has S=110.6mJy.
\end{table*}

\subsection{The new sample of $z>3$ quasars}

\begin{figure*}
\centerline{
\psfig{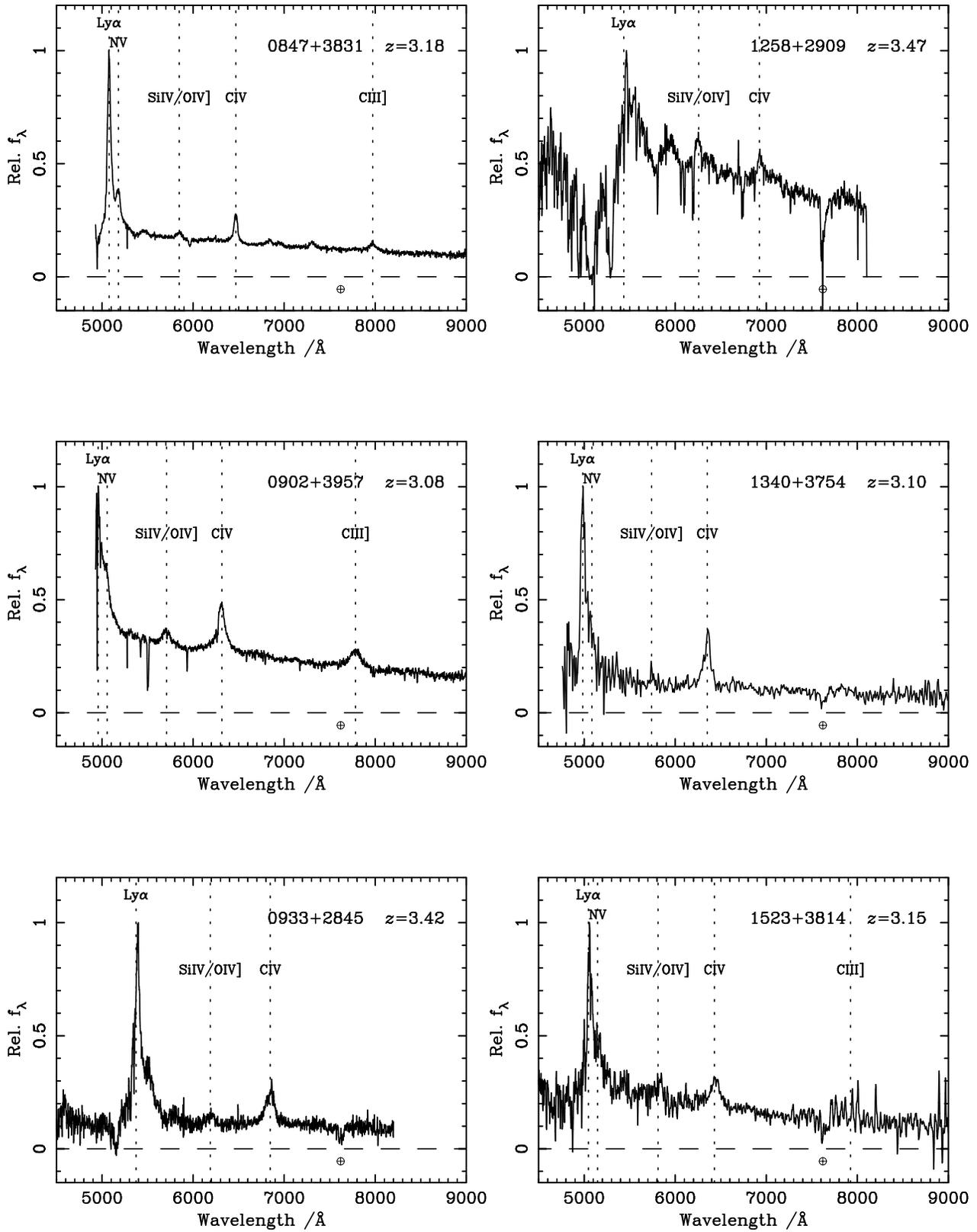}}
\caption[]{Spectra of the $z>3$ quasars in the GB/FIRST flat-spectrum sample to date.}
\label{spec}
\end{figure*}

\begin{figure*}
\centerline{
\psfig{figure=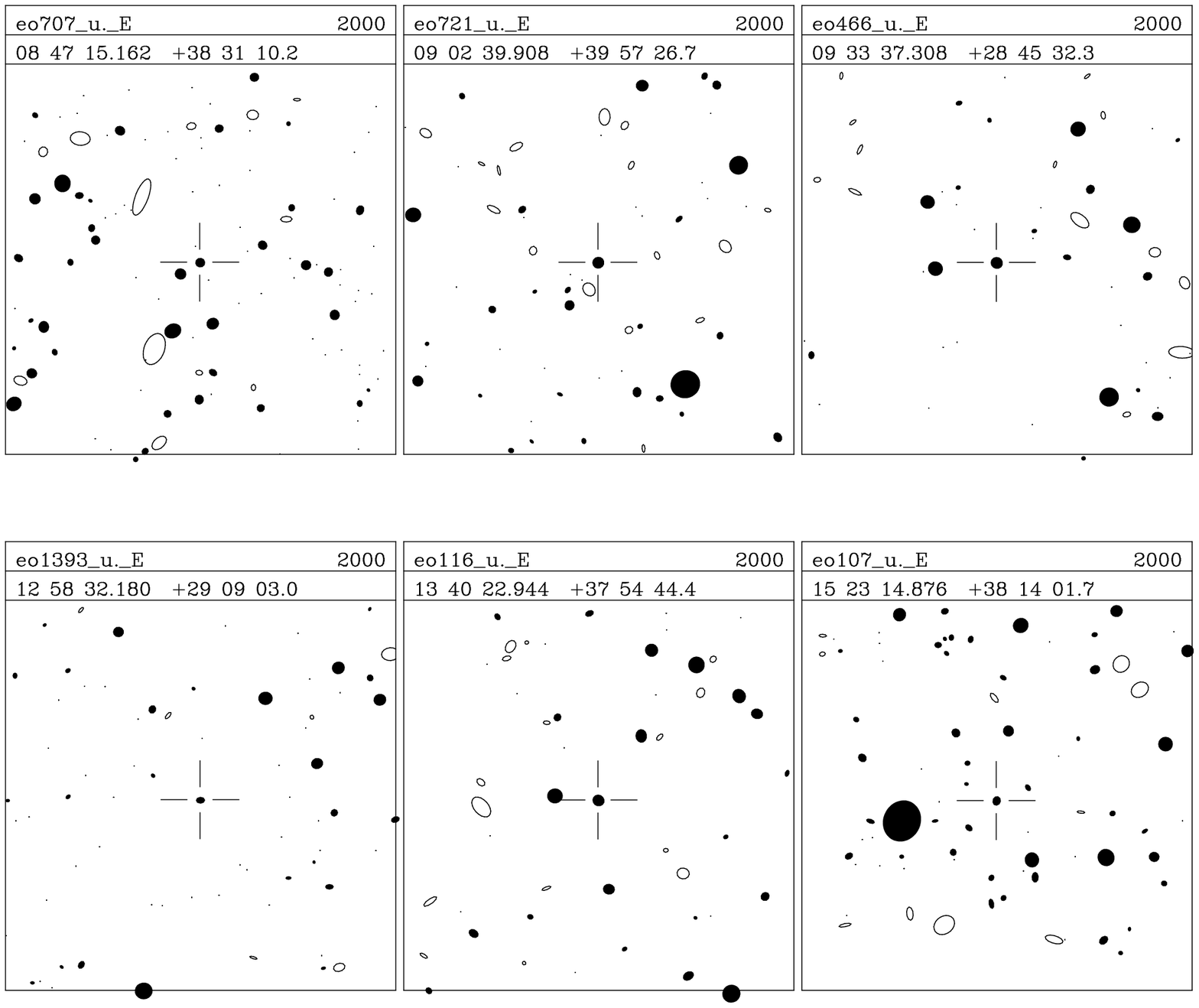,height=5.0in,bbllx=24pt,bblly=284pt,bburx=588pt,bbury=757pt}
}
\caption[]{Finding charts from APM scans of POSS-I E
plates ($\rm \sim R$ band) for the $z>3$ quasars in the GB/FIRST
flat-spectrum sample
to date. North is at the top, East to the left. The central cross is 1
arcmin across.}
\label{fct}
\end{figure*}

A summary of the six $z>3$ quasars in the sample to date is given in
Table~4, their spectra are shown in Figure~\ref{spec} and finding
charts are given in Figure~\ref{fct}. Note that the spectra obtained
at Keck (0847+3831 and 0902+3957) have been corrected for atmospheric
absorption bands while the others have not.  The spectrum of 0933+2845
($z=3.42$, Gregg et al 1996) was obtained at Lick on 1995 July 3 as
part of the FBQS. 1258+2909 was re-observed at Lick 3m on 1996 March
15 to obtain a higher signal-to-noise spectrum.  This spectrum, rather
than the discovery spectrum, is shown in Figure~\ref{spec}.

\begin{table*}
\caption{Summary of the $z>3$ QSOs in the sample to date.}
\begin{tabular}{llrcclrrl}\hline
  $\rm Name^1$   &\multicolumn{1}{c}{$z$}   & $N\sigma_c$ & E & $\rm O-E$ & $\Delta r/''$
& $\rm S_{GB}$/mJy  & $\rm \alpha_{1.4GHz}^{5GHz}$ & Comments\cr \hline
 0847+3831 &  3.18 &$-1.01$& 17.55  & 1.37  &   0.31  &  124  &$-0.32$& Keck 03/97                        \cr
 0902+3957 &  3.08 & 0.83 & 17.56  & 1.24  &   0.14  &   31  &$-0.03$& Keck 03/97                        \cr
 0933+2845 &  3.42 & 0.27 & 17.50  & 1.27  &   0.16  &   66  &$-0.48$&
Previously known, 
 Gregg et al 1996\cr
 1258+2909 &  3.47 & 0.32 & 18.52  & 2.06  &   0.48  &   32  &$0.19$& Lick 04/95  \cr
 1340+3754 &  3.10 & 0.08 & 17.85  & 1.68  &   0.85  &  305  &$0.06$&
 Previously known, Hook et al 1995 \cr
 1523+3814 &  3.15 &$-1.09$& 18.38  & 1.40  &   0.40  &   27 &$-0.45$& Lick 05/96  \cr\hline\hline
\end{tabular}
\label{z3tab}
\raggedright

\noindent{\bf Notes}\newline
1 based on J2000 position \newline
\end{table*}

Two $z>3$ radio-loud quasars are known to lie within the survey area
but were not found by this survey. FBQS J1021+3001 with $z=3.1$ (Gregg
et al 1996) is in the FIRST catalogue but is below the flux limit used
in this search.  B3 1239+376 with $z=3.82$ (Vermeulen et al. 1996) is
optically fainter than the limit of the APM scans.

\section{Conclusions}
In this paper we have described a survey for high redshift radio-loud
quasars with a limiting flux density of 25mJy, about a factor of 10
fainter than previous surveys. This survey has become feasible because
of new accurate positions for large numbers of faint radio sources
supplied by the VLA FIRST survey. Six $z>3$ quasars have so far been
found in an area of 0.49sr ($\sim 1610$ sq deg), after completing
follow up for $70\%$ of the spectroscopic sample. The effective area
covered so far is thus $70\% \times 1610 = 1130$ square
degrees. Therefore we find a surface density of flat-spectrum $z>3$
quasars of one per 190 square degrees to a radio flux limit of $\rm
S_{5GHz}\ge 25$mJy and an optical limit of E=19.5mag.  After
completion of the spectroscopic observations the new sample will be
used in an analysis of the space density of quasars at high redshift.

\section*{ACKNOWLEDGMENTS} IMH acknowledges a PPARC/NATO fellowship and
RGM thanks the Royal Society for support.  This research has made use
of the NASA/IPAC Extragalactic Database (NED) which is operated by the
Jet Propulsion Laboratory, Caltech, under contract with the National
Aeronautics and Space Administration.  Some observations described in
this paper were obtained at the Keck-II 10m telescope of the
California Association for Research in Astronomy.  We thank Daniel Stern,
Lisa Storrie-Lombardi, Robert Knop and Saul Perlmutter for assistance
with the observations.

\end{document}